\documentclass[aps,prl,twocolumn,superscriptaddress]{revtex4-2}
\usepackage{graphicx}
\usepackage{amsmath}
\usepackage{amssymb}
\usepackage{colordvi}
\usepackage{mathrsfs}
\usepackage{bm}
\usepackage{verbatim}
\usepackage{dcolumn}
\usepackage{epsfig}
\usepackage{subfigure}
\usepackage{makecell}
\usepackage[colorlinks,linkcolor=blue,anchorcolor=blue,citecolor=blue]{hyperref}

\begin{document}
\title{Chern Number Tunable Quantum Anomalous Hall Effect in Compensated Antiferromagnets}

\author{Wenhao Liang}
\thanks{These authors contributed equally to this work.}
\affiliation{International Centre for Quantum Design of Functional Materials, CAS Key Laboratory of Strongly-Coupled Quantum Matter Physics, and Department of Physics, University of Science and Technology of China, Hefei, Anhui 230026, China}

\author{Zeyu Li} 
\thanks{These authors contributed equally to this work.}
\affiliation{International Centre for Quantum Design of Functional Materials, CAS Key Laboratory of Strongly-Coupled Quantum Matter Physics, and Department of Physics, University of Science and Technology of China, Hefei, Anhui 230026, China}

\author{Jiaqi An}
\affiliation{International Centre for Quantum Design of Functional Materials, CAS Key Laboratory of Strongly-Coupled Quantum Matter Physics, and Department of Physics, University of Science and Technology of China, Hefei, Anhui 230026, China}

\author{Yafei Ren} \email[Correspondence author:~~]{yfren@udel.edu}
\affiliation{Department of Physics and Astronomy, University of Delaware, Newark, Delaware 19716, USA}

\author{Zhenhua Qiao} \email[Correspondence author:~~]{qiao@ustc.edu.cn}
\affiliation{International Centre for Quantum Design of Functional Materials, CAS Key Laboratory of Strongly-Coupled Quantum Matter Physics, and Department of Physics, University of Science and Technology of China, Hefei, Anhui 230026, China}
\affiliation{Hefei National Laboratory, University of Science and Technology of China, Hefei 230088, China}

\author{Qian Niu}
\affiliation{International Centre for Quantum Design of Functional Materials, CAS Key Laboratory of Strongly-Coupled Quantum Matter Physics, and Department of Physics, University of Science and Technology of China, Hefei, Anhui 230026, China}

\date{\today{}}
	
\begin{abstract} 
\textcolor{blue}{We propose to realize the quantum anomalous Hall effect (QAHE) in two-dimensional compensated antiferromagnets without net spin magnetization.} We consider antiferromagnetic MnBi$_2$Te$_4$ as a concrete example. \textcolor{blue}{By breaking the parity-time ($\mathcal{PT}$) symmetry of even-layer MnBi$_2$Te$_4$, we find that the system can host the QAHE with a nonzero Chern number.} We show that by controlling the antiferromagnetic spin configuration, for example, down/up/up/down that breaks $\mathcal{PT}$ symmetry, tetralayer MnBi$_2$Te$_4$ can host a Chern number $\mathcal{C}=-1$. Such spin configuration can be stabilized by pinning the spin orientations of the surfaces. \textcolor{blue}{Furthermore, via tuning the on-site orbital energy and vertical electric fields, we find rich QAHE phases with tunable Chern number of $|\mathcal{C}|=1, 2, 3$.
In addition, we reveal that the edge states are layer-selective and primarily locate at the boundaries of the bottom and top layers.}
Our work not only proposes a scheme to realize Chern number tunable QAHE in antiferromagnets without net spin magnetization, but also provides a platform for layer-selective dissipationless transport devices.
\end{abstract}

\maketitle

The quantum anomalous Hall effect (QAHE) exhibits topologically protected chiral edge states, the dissipationless feature of which makes them attractive for next-generation high-performance electronics~\cite{review-1}. The QAHE also shows a close connection with novel quantum phenomena such as topological magnetoelectric effects and topological superconductivity~\cite{review-2, axion, tp-sc}. The search for QAHE is thus a hot spot in condensed matter physics~\cite{RMP-1, RMP-2, RMP-3} with many recipes being theoretically proposed~\cite{QAHE-1, QAHE-2, QAHE-3, QAHE-4, QAHE-5, QAHE-6, QAHE-7, QAHE-8, QAHE-9, QAHE-10, QAHE-11, QAHE-12, QAHE-13, QAHE-14, liufeng}. In experiments, the QAHE has been observed in three main categories, i.e., magnetic doped topological insulators~\cite{QAHE-15}, intrinsic magnetic topological insulators~\cite{QAHE-16} and moir$\acute{\text{e}}$ systems~\cite{QAHE-17,QAHE-18}. All of them possess ferromagnetism, which can be influenced by fluctuations of external magnetic fields that can arise from stray fields or other external sources, which is not desired in applications. 
In contrast, antiferromagnets are more robust to fluctuations of magnetic fields~\cite{AFM-review-1} and have attracted growing attention in recent years, which extended the traditional spintronics to antiferromagnetic spintronics~\cite{AFM-review-1,AFM-review-2,AFM-review-3,AFM-review-4,AFM-review-5,AFM-1,AFM-2,AFM-3,AFM-4}. The realization of QAHE in antiferromagnets is thus highly desired for applications. Despite several theoretical proposals of realizing QAHE in antiferromagnetic systems~\cite{AFM-QAHE-1,AFM-QAHE-2,AFM-QAHE-3,AFM-QAHE-4,AFM-QAHE-5,AFM-QAHE-6,AFM-QAHE-7,AFM-QAHE-8,AFM-QAHE-Lei,AFM-QAHE-9,AFM-QAHE-10}, it is still challenging to realize Chern number tunable QAHE in compensated antiferromagnets without net spin magnetization.

\begin{figure}
	\includegraphics[width=7.5cm,angle=0]{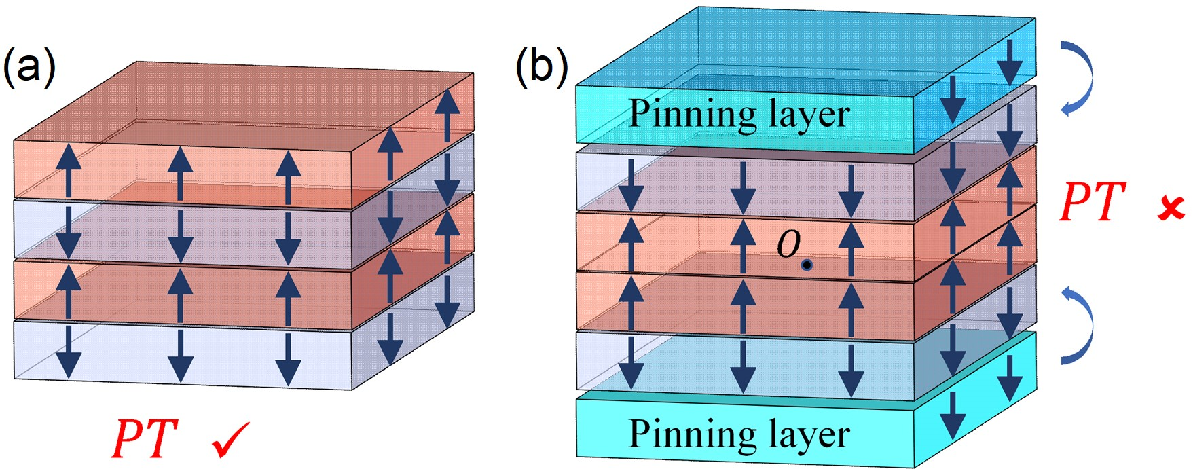}
	\caption{\textcolor{blue}{The tetralayer (a) intrinsic A-type antiferromagnets with $\mathcal{PT}$ symmetry, and (b) pinned antiferromagnets without $\mathcal{PT}$ symmetry.} The space inversion point is denoted by $O$.}
	\label{structure_4layers}
\end{figure}

\begin{figure*}
	\includegraphics[width=15cm,angle=0]{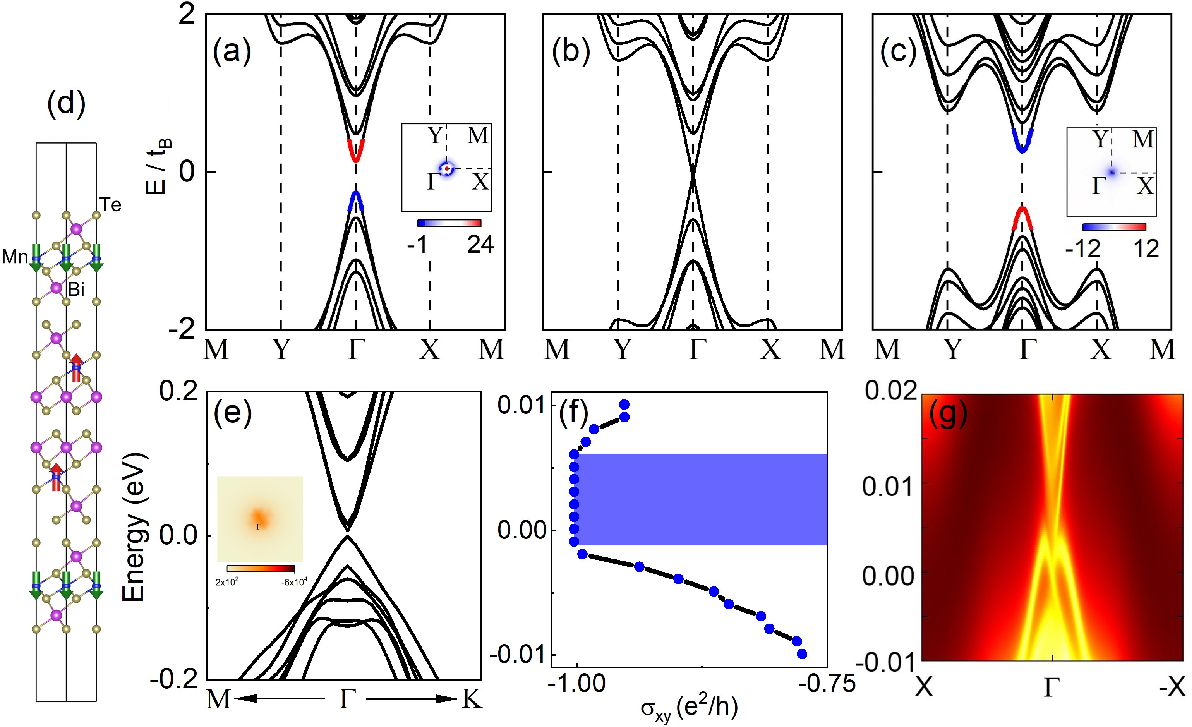}
	\caption{Bulk band structures along the high symmetry lines of tetralayer antiferromagnets without $\mathcal{PT}$ symmetry for $t_A=$ (a) $-1.5$ (b) $-1.38$ (c) $-1$. The majority of the electronic spin is up (red) or down (blue) near the Fermi level. The inset is the distribution of Berry curvatures in the momentum space, with black lines marking the first Brillouin zone. \textcolor{blue}{(d) Side view of four-septuple-layer MnBi$_2$Te$_4$ film, with magnetic configuration represented by the red and green arrows. First-principle calculations for (e) Bulk band structure with spin-orbit coupling, where the inset is the distribution of Berry curvatures around $\Gamma$ point. (f) The anomalous Hall conductivity. (g) Energy spectra with edge state of the semi-infinite ribbon. }}
	\label{bulk1}
\end{figure*}

In this Letter, we propose to realize the antiferromagnetic QAHE in an experimentally feasible system, i.e., the magnetic topological insulator, by controlling the spin configuration. 
\textcolor{blue}{We consider the tetralayer MnBi$_2$Te$_4$~\cite{ MBT-0, MBT-01, MBT-1,MBT-2,MBT-3,MBT-4,MBT-5,layer Hall,MBT-6,MBT-7,MBT-8,MBT-9,MBT-10,MBT-11} as an example, which is an intrinsic antiferromagnetic topological insulator and has attracted broad interest as it bridges the topology, magnetism, and van der Waals materials. The recent advancements in sample quality and material control capability have revitalized the field~\cite{exp-1, exp-2,exp-3}.
This material typically shows A-type antiferromagnetic spin configuration (e.g., down/up/down/up), which is invariant under $\mathcal{PT}$ operation where $\mathcal{P}$ is parity and $\mathcal{T}$ is the time-reversal operation as illustrated in Fig.~\ref{structure_4layers}(a). This symmetry guarantees a zero Chern number.
By introducing magnetic pinning layers that fix the spin configuration of the surfaces, the stable spin configuration deviates from the A-type one as illustrated in Fig.~\ref{structure_4layers}(b) where the spin orientations in the middle layers are the same while they are opposite to the spin orientations in the top and bottom layers. This new spin configuration breaks $\mathcal{PT}$ symmetry as it is even under $\mathcal{P}$ whereas odd under $\mathcal{T}$. The breaking of both $\mathcal{PT}$ and $\mathcal{T}$ symmetries allows a nonzero Chern number.
Our calculation confirms that this tetralayer MnBi$_2$Te$_4$ indeed hosts QAHE with $\mathcal{C}=-1$ by both model and first-principle calculations.
Our results show that the stacking of two bilayer systems with zero Chern number leads to a nonzero Chern number, which is quite surprising and unusual. 
Furthermore, by tuning the on-site orbital energy and the electric fields, the tetralayer antiferromagnets can have rich QAHE phases with tunable Chern numbers, i.e., $|\mathcal{C}|=1, 2, 3$. 
The edge states of the QAHE are primarily distributed at the boundaries of the surfaces.}

\emph{Antiferromagnetic QAHE}.--- \textcolor{blue}{The antiferromagnetic topological insulators can be described by the model Hamiltonian on a layered square lattice}~\cite{MBT-7,MBT-8,MBT-9,MBT-10,MBT-11,jianghua,BHZ,chengran} 
\begin{eqnarray}\label{model hamiltonian}
	H = \sum_{i} c_i^\dagger (E_0+\nu_{i} m)  c_i + \sum_{\left\langle ij\right\rangle, \alpha}c_i^\dagger T_\alpha c_j + \text{H.c.},
\end{eqnarray}
where $c_i=[c_{+\uparrow},c_{-\uparrow},c_{+\downarrow},c_{-\downarrow}]^T$
are the annihilation operators of electronic states at site $i$ where $\pm$ represent two different orbits and $(\uparrow,\downarrow)$ spin indices. $E_0 = (t_A\sigma_0 \otimes \tau_z - \frac{3}{2}A \sigma_0 \otimes \tau_0)$, $T_{\alpha} =\frac{1}{2}(t_B \sigma_0 \otimes \tau_z + A \sigma_0 \otimes \tau_0 - \text{i}B \sigma_\alpha \otimes \tau_x)$, with $\alpha=x, y, z$. 
The Pauli matrices $\bm{\sigma}$ and $\bm{\tau}$ are for spin and orbit degrees of freedom, respectively.
$\left\langle ij \right\rangle$ denotes the nearest neighboring coupling. $m =m_0\sigma_z \otimes \tau_0$ is the exchange field and $\nu_{i}=\{-1,1,1,-1\}$ representing down/up/up/down spin configuration. 
\textcolor{blue}{The spin-orbit coupling parameter $B$ reflects the Fermi velocity, $t_A$ is the on-site orbital energy that determines the inverted band gap, and it can be controlled by atomic doping~\cite{doping} or pressure~\cite{pressure}. Unless otherwise noted, we set the other parameters to be $A = 0.1, B = 1.5, m_0=0.35$, and $t_B = 1$~\cite{jianghua}.}

\begin{figure*}
	\includegraphics[width=14cm,angle=0]{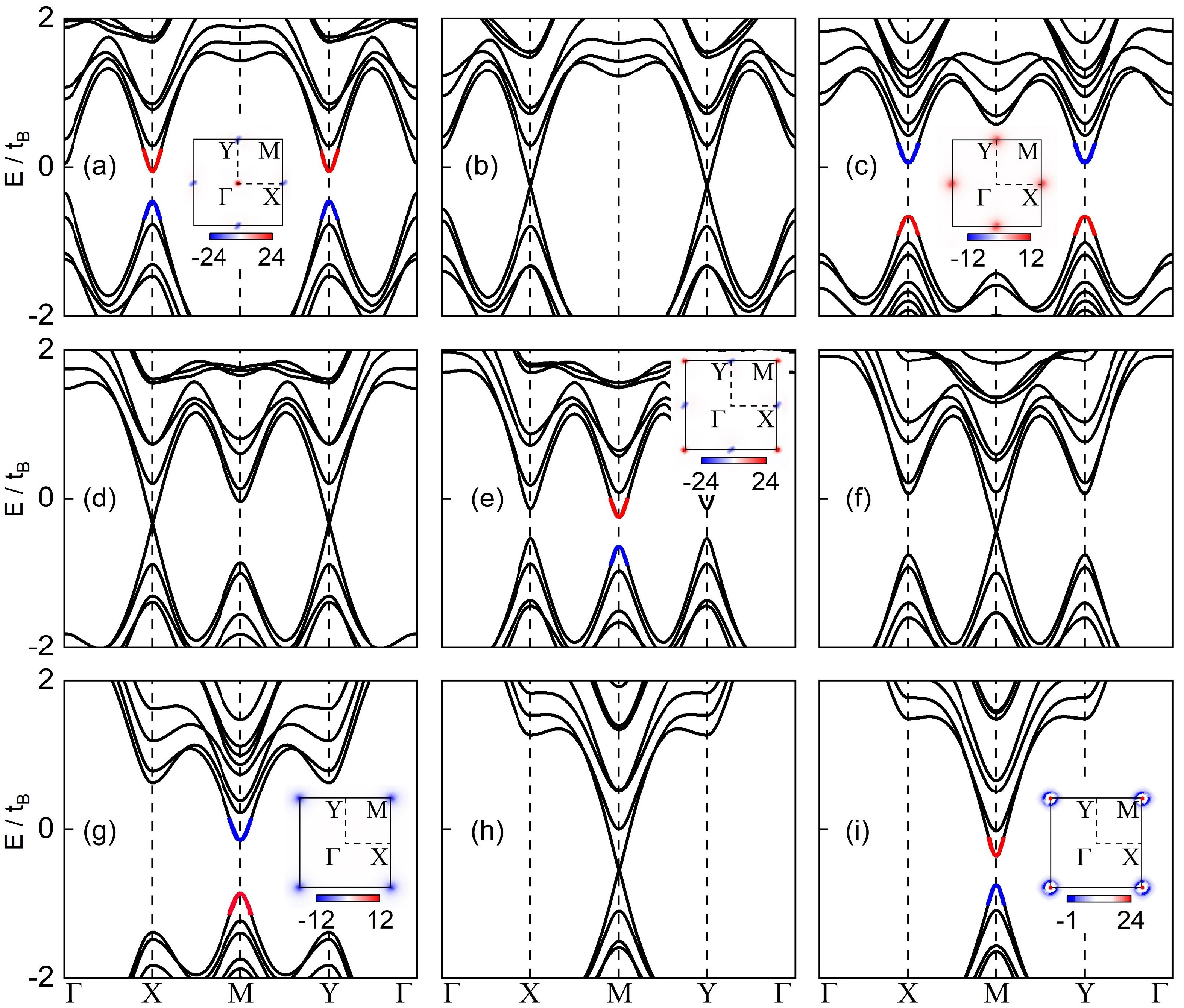}
	\caption{The evolution of bulk band structures for different $t_A$: (a) $-0.5$ (b) $-0.38$ (c) $0$ (d) $0.38$ (e) $0.5$ (f) $0.62$ (g) $1$ (h) $1.38$ (i) $1.5$. The majority of the electronic spin is up (red) or down (blue) near the Fermi level. The inset is the distribution of Berry curvatures in the momentum space, with black lines marking the first Brillouin zone.}
	\label{bulk2}
\end{figure*}

The spin-compensated antiferromagnetic topological insulator with a down/up/up/down spin configuration can host a Chern insulator phase. 
In Fig.~\ref{bulk1}, we plot the band structures with different $t_A$ where the band structures are non-degenerate since the $\mathcal{PT}$ symmetry is broken. A topological phase transition happens as $t_A$ changes. For $t_A=-1.5$, as shown in Fig.~\ref{bulk1}(a), the valance and conductance band near the Fermi level are dominated by spin-down and spin-up electrons, respectively. The band gap is topologically trivial as the positive and negative Berry curvatures (see the inset) cancel in the first Brillouin zone resulting in a vanished Chern number. 
The band gap decreases with increasing $t_A$, and at $t_A=-1.38$, a Dirac cone emerges around $\Gamma$ point [see Fig.~\ref{bulk1}(b)]. When $t_A$ exceeds this value, the band gap reopens, accompanied by the switch of spin between the valance and conductance band near the Fermi level [see Fig.~\ref{bulk1}(c)]. 
In this case, only negative Berry curvatures emerge concentrated at the $\Gamma$ point, leading to $\mathcal{C}=-1$, i.e., the system goes into the QAHE states. 

\textcolor{blue}{Our first-principles calculation confirmed that our models and parameters can faithfully describe the physics of MnBi$_2$Te$_4$ near $\Gamma$ point. Figure~\ref{bulk1}(d) illustrates the lattice and spin structure for our first-principle calculations of tetralayer MnBi$_2$Te$_4$ with down/up/up/down spin configuration (see details in Supplementary Materials~\cite{supple}). 
The electronic structure is shown in Fig.~\ref{bulk1}(e) where a global band gap exists. The inset shows Berry curvature that concentrates around the $\Gamma$ point. We further calculated the intrinsic anomalous Hall conductivity at different Fermi energy in Fig.~\ref{bulk1}(f) where we find quantized anomalous Hall conductivity $\sigma_{xy}=-e^2/h$ when the Fermi energy lies in the band gap as highlighted by the shaded region. We further calculated the spectrum function for a semi-infinite slab using the iterative Green's function method as shown in Fig.~\ref{bulk1}(g), which displays the chiral edge states.
Therefore, the first-principle calculations prove the reliability of our theoretical model and results, i.e., the QAHE can be realized in compensated antiferromagnets MnBi$_2$Te$_4$.}

\emph{Phase diagram.---} \textcolor{blue}{To reveal the robustness of the QAHE against tuning parameters and identify possible topological phases that can be hosted by the spin-compensated antiferromagnets, we systematically study the parameter space. We first study the effect of $t_A$. By increasing $t_A$, the band structures are notably deformed as shown in Fig.~\ref{bulk2}. 
When $t_A=-0.38$, two Dirac points are formed, and the position migrates from $\Gamma$ point to $\text{X}$/$\text{Y}$ points. Then the band inversion occurs again, i.e., another topological phase transition emerges.}
As shown in Fig.~\ref{bulk2}(c), only positive Berry curvatures are observed at $\text{X}$/$\text{Y}$ points. Since the Berry curvature around each $\text{X}$/$\text{Y}$ point contributes 1 to the total Chern number, the system enters the QAHE state with $\mathcal{C}=2$. 
With the further increase of $t_A$, the position of the band gap continues to change. When $t_A=0.62$, the Dirac point changes to the $\text{M}$ point.
After $t_A$ exceeds this value, the band gap opens and accompanies band inversion at $\text{M}$ point [see Fig.~\ref{bulk2}(g)]. Accordingly, the system possesses negative Berry curvatures concentrated at $\text{M}$ point, which induces a Chern insulator with $\mathcal{C}=-1$. 
During the whole process, the system undergoes topological phase transitions six times. The band gap migrates from $\Gamma$ point to $\text{X}/\text{Y}$ points, and subsequently to $\text{M}$ point, which induces the redistribution of Berry curvatures leading to a tunable Chern number ranging from $-1$ to $2$ and then to $-1$.
It is noted that the two QAHE with $\mathcal{C}=-1$ have different band structures and distributions of the Berry curvatures.

\begin{figure}
	\includegraphics[width=8cm,angle=0]{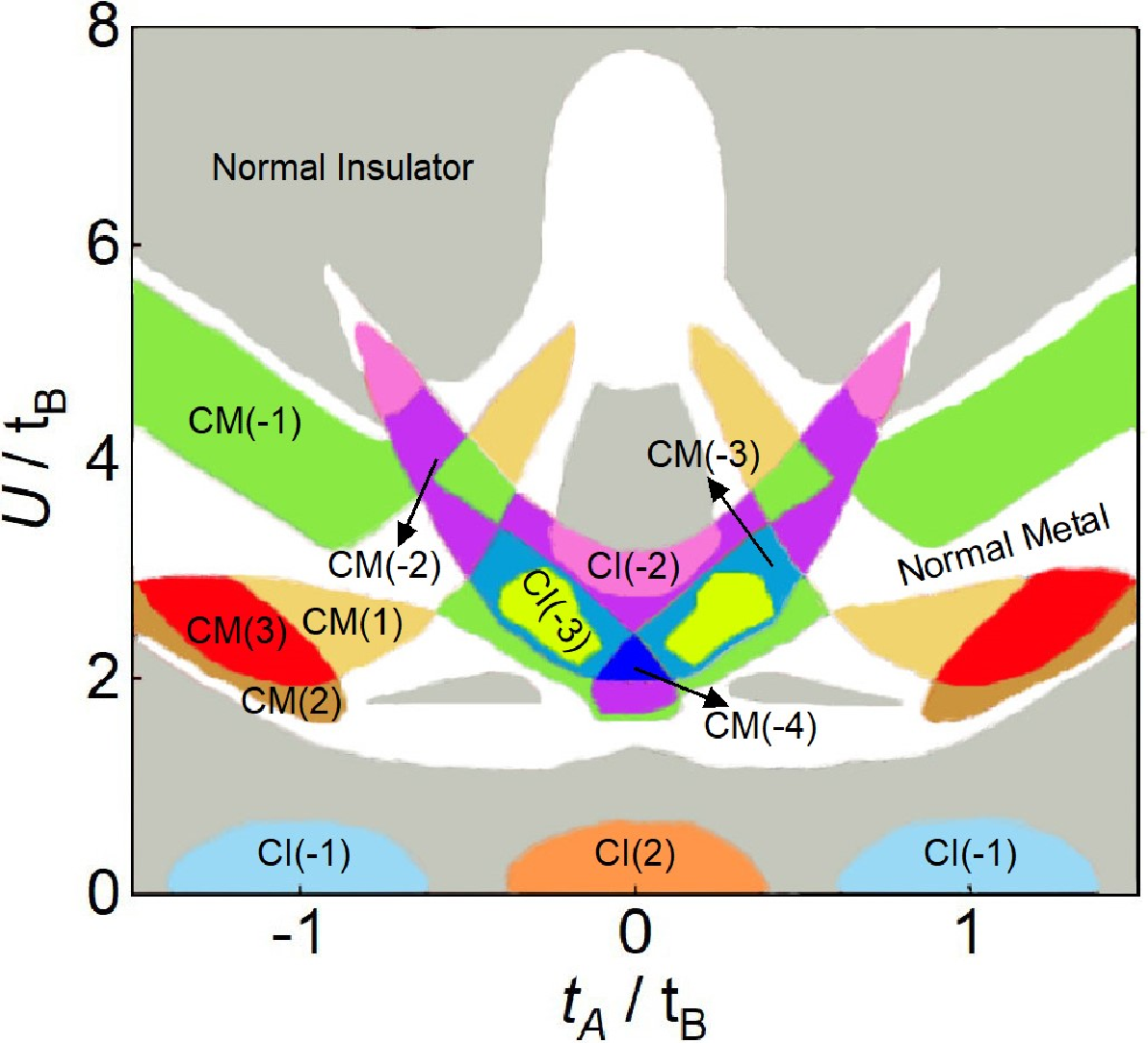}
	\caption{ \textcolor{blue}{ The phase diagram of Chern number (the number in brackets) as a function of $t_A$ and $U$. The different phases are Chern insulator (CI), Chern metal (CM), normal insulator, and normal metal. }}
	\label{phase}
\end{figure}

\textcolor{blue}{Furthermore, the vertical electric field is another way to tune the topology in antiferromagnetic topological insulators~\cite{Efield-1,Efield-2}. To investigate the influence of the electric fields on the topological properties, we add the gate voltage $H_E= \sum_i  U_i c_i^{\dagger} c_i$, where $U_i$ measures the potential strength at layer $i$. In tetralayer, the potential is $\left\lbrace U,U/3,-U/3,-U\right\rbrace $ for each layer respectively. The phase diagram is plotted in Fig.~\ref{phase} where we find that both $t_A$ and $U$ can effectively tune the topological properties leading to rich topological phases. These phases include Chern insulators, characterized by a global band gap and a nonzero Chern number, as well as Chern metals, where the low-energy bands exhibit local gaps and carry a nonzero total Chern number, while bands near the Fermi level cross it~\cite{ChernMetal}. In the presence of weak disorder, the small Fermi pockets may become localized, leading to Anderson insulators with nontrivial topology. More details can be found in the Supplemental Materials~\cite{supple}.}

\begin{figure}
  \includegraphics[width=8.5cm,angle=0]{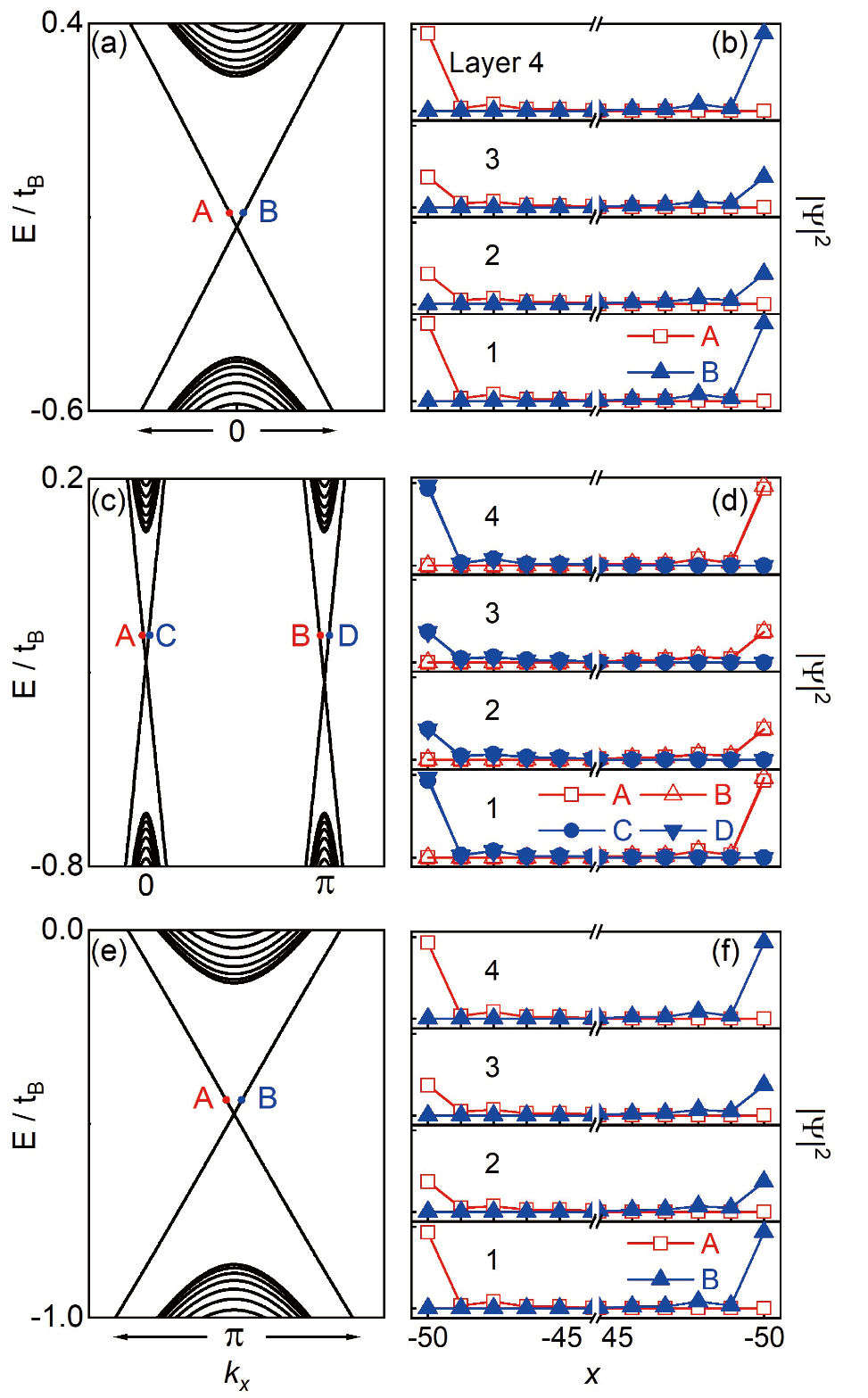}
  \caption{The left column: the one-dimensional energy spectrum of QAHE for (a) $t_A=-1$, (c) $t_A=0$, and (e) $t_A=1$, where the different edge states near the charge neutral point are labeled as $``A", ``B", ``C"$ and $``D"$. The right column: the corresponding wavefunction distributions $\left|\psi\right| ^2$ of different edge states, with values ranging from 0 to 0.32 in each layer. A break is taken on the $x$-axis since the values within the break range are zero. }
  \label{ribbons}
\end{figure}

\emph{Layer-selective edge states.---} As a multi-layer system, the MnBi$_2$Te$_4$ possesses the spatial degree of freedom corresponding to different layers, which will bring new features to electronic and topological properties~\cite{MBT-10,layer Hall}. 
\textcolor{blue}{Here we take the QAHE without an electric field as an example to reveal the layer-selective characteristics of edge states. The $U=0$ if not specifically mentioned in this work. }Figure~\ref{ribbons}(a) displays the one-dimensional band structure for $t_A=-1$ with $\mathcal{C}=-1$. There is one pair of gapless chiral edge states at $k_x=0$, whose wavefunction distributions are depicted in Fig.~\ref{ribbons}(b). The state $A$ with negative velocity is located at the left boundary. It is noted that the states are not equally divided in each layer. The wavefunctions concentrate on the surfaces, i.e., the top and bottom layers, with a much larger probability than that in the middle layers. The case is similar for state B, which is the inversion counterpart of state A localized on the opposite boundary with opposite velocity.

The layer selectivity is similar for QAHE with different Chern numbers. With $t_A=0$, the system is a Chern insulator with $\mathcal{C}=2$. According to bulk-boundary correspondence~\cite{bulk-boundary}, there are two pairs of gapless chiral edge states, appearing at $k_x=0$ and $k_x=\pi$ respectively, as shown in Fig.~\ref{ribbons}(c). The wavefunction distributions [see Fig.~\ref{ribbons}(d)] indicate that the edge states are also primarily distributed on the surfaces. 
Similarly, the system for $t_A=1$ is a Chern insulator with $\mathcal{C}=-1$, and the corresponding chirality is identical to that of $t_A=-1$. Its chiral edge states at $k_x=\pi$ [see Fig.~\ref{ribbons}(e)] are also primarily distributed on the surfaces, as shown in Fig.~\ref{ribbons}(f). 
Such layer-selective phenomena of these QAHE can be applied in layer-selective dissipationless transport devices. This behavior is also illustrated by the local density of states~\cite{supple}. 
\textcolor{blue}{In addition, our QAHE is robust against weak and moderate magnetic disorder~\cite{supple}.}

\emph{Stability of the spin configuration.---} Although the spin configuration in our proposal is not the lowest-energy state compared to the intrinsic A-type antiferromagnetic configuration, it can be achieved by pinning the top and the bottom layers, \textcolor{blue}{which is widely used in magnetic heterostructures and devices~\cite{pinning, pinning-1,pinning-2,pinning-3,pinning-4}.
Fortunately, the magnetic pinning layer has little influence on electronic band structures and topological properties of the studied system~\cite{pinning-3,pinning-4}. Here we show that, when we pin the spin configurations of the top and bottom layers downward, the down/up/up/down spin configuration [Fig.~\ref{structure_4layers}(b)] is the lowest energy configuration.}
We consider a Heisenberg model to describe the spin configuration,
\begin{equation}
	H^s=J_{\mathrm{c}} \sum_{\langle i j\rangle} \bm{S}_i \cdot \bm{S}_j-D \sum_i\left(\bm{S}_i\cdot\hat{\bm{z}}\right)^2,\label{eq_heisenberg}
\end{equation}
where $\bm{S}_i$ labels the spin at layer $i$ 
\begin{equation}
	\bm{S}_i=\{\sin(\theta_i)\cos(\phi_i),\sin(\theta_i)\sin(\phi_i),\cos(\theta_i)\},\label{eq_spin}
\end{equation}
with $\theta$ and $\phi$ being the spin polar and azimuthal angles, respectively. 
$J_{\mathrm{c}}>0$ is the interlayer magnetic exchange interaction between the nearest-neighbor layers $\langle i j\rangle$, $D$ is the uniaxial magnetic anisotropy. 
The pinning layer fixes the spin orientations at the top and bottom layers downward with $\theta_1=\theta_4=\pi$. The spin orientations of
the middle two layers are determined by the lowest energy
configuration. \textcolor{blue}{Given the parameters $J_c=0.034$ meV and $D=0.03$ meV in MnBi$_2$Te$_4$~\cite{heisenberg}, plugging Eq.~\eqref{eq_spin} into Eq.~\eqref{eq_heisenberg} and minimizing the total energy as a function of the polar and azimuthal angles, we get the spin configuration of the lowest energy at $\theta_2=\theta_3=0$. Our first-principle calculations obtain the same results using CrI$_3$ as pinning layers~\cite{supple}.} Therefore, the spin configuration of down/up/up/down we studied is stable when we pin the spin orientations of the top and bottom layers. 

\emph{Summary.---} We systematically investigate the electronic and topological properties of two-dimensional compensated antiferromagnets. \textcolor{blue}{Taking even-layer MnBi$_2$Te$_4$ as an example, the QAHE can be realized by controlling the spin configurations through magnetic pinning and thus breaking the $\mathcal{PT}$ symmetry.
Furthermore, the distribution of Berry curvatures is successively tuned via tuning the on-site orbital energy or the electric gate voltage, giving rise to tunable Chern numbers and rich QAHE phases with $|\mathcal{C}|=1, 2, 3$.} We also find that the edge states are layer-selective, i.e., primarily distributed at the boundaries of the surfaces. Our work not only provides an ideal platform to realize Chern number tunable QAHE in compensated antiferromagnets that have no net spin magnetization, but also sheds light on layer-selective dissipationless transport for practical applications.
\textcolor{blue}{It is also noted that in multi-layer systems, for example, hexalayer systems, the Chern number tunable QAHE also emerges~\cite{supple}.}

\begin{acknowledgements}
We are grateful to Prof. Feng Liu for valuable discussions. W. L, J.A. and Z.Q. are supported by the National Natural Science Foundation of China (Grants No. 11974327 and No. 12004369), Natural Science Basic Research Program of Shanxi (No. 20210302124252), Anhui Initiative in Quantum Information Technologies (AHY170000), and Innovation Program for Quantum Science and Technology (2021ZD0302800). Z.L. is supported by China Postdoctoral Science Foundation (2023M733411 and 2023TQ0347). Y.R. acknowledges startup funds provided by the College of Arts and Sciences and the Department of Physics and Astronomy of the University of Delaware. Q. N. is supported by the National Natural Science Foundation of China (Grant No. 12234017). We also thank the Supercomputing Center of University of Science and Technology of China for providing the high performance computing resources.
\end{acknowledgements}


\begin{thebibliography}{99}
%\bibitem{QHE}
%K. V. Klitzing, G. Dorda, and M. Pepper, New Method for High-Accuracy Determination of the Fine-Structure Constant Based on Quantized Hall Resistance, \text{Phys. Rev. Lett.} \textbf{45}, 494 (1980).
	
%\bibitem{Haldane}
%F. D. M. Haldane, Model for a Quantum Hall Effect without Landau Levels: Condensed-Matter Realization of the ``Parity Anomal'', \text{Phys. Rev. Lett.} \textbf{61}, 2015 (1988).
	
\bibitem{review-1}
H. Weng, R. Yu, X. Hu, X. Dai and Z. Fang, Quantum anomalous Hall effect and related topological electronic states, \text{Adv. Phys.} \textbf{64}, 227-282 (2015).
	
\bibitem{review-2}
K. He, Y. Wang, and Q.-K. Xue, Topological Materials: Quantum Anomalous Hall System, \text{Annu. Rev. Condens. Matter Phys.} \textbf{9}, 329 (2018).
	
\bibitem{axion}
D. Xiao, J. Jiang, J.-H. Shin, W. Wang, F. Wang, Y.-F. Zhao, C. Liu, W. Wu, M. H. W. Chan, N. Samarth et al., Chang, Realization of the Axion Insulator State in Quantum Anomalous Hall Sandwich Heterostructures, \text{Phys. Rev. Lett.}   \textbf{120}, 056801 (2018).
	
\bibitem{tp-sc}
X.-L.Qi, T. L. Hughes, and S.-C. Zhang, Chiral topological superconductor from the quantum Hall state, \text{Phys. Rev. B}   \textbf{82}, 184516 (2010).
	
\bibitem{RMP-1}
C.-Z. Chang, C.-X. Liu, and A. H. MacDonald, Colloquium: Quantum anomalous Hall effect, \text{Rev. Mod. Phys.} \textbf{95}, 011002 (1988).
	
\bibitem{RMP-2}
X.-L. Qi and S.-C. Zhang, Topological insulators and superconductors, \text{Rev. Mod. Phys.} \textbf{83}, 1057 (2011).
	
\bibitem{RMP-3}
M. Z. Hasan and C. L. Kane, Colloquium: Topological insulators, \text{Rev. Mod. Phys.} \textbf{82}, 3045 (2010).
	
	
\bibitem{QAHE-1}
C.-X. Liu, X.-L. Qi, X. Dai, Z. Fang, and S.-C. Zhang, Quantum Anomalous Hall Effect in Hg$_{1-y}$Mn$_y$Te Quantum Wells, \text{Phys. Rev. Lett.}  \textbf{101}, 146802 (2008).
	
\bibitem{QAHE-2}
R. Yu, W. Zhang, H.-J. Zhang, S.-C. Zhang, X. Dai, and Z. Fang, Quantized Anomalous Hall Effect in Magnetic Topological Insulators, \text{Science}   \textbf{329}, 61 (2010).
	
\bibitem{QAHE-3}
Z. Qiao, S. A. Yang, W.-X. Feng, W.-K. Tse, J. Ding, Y. G. Yao, J. Wang, and Q. Niu, Quantum anomalous Hall effect in graphene from Rashba and exchange effects, \text{Phys. Rev. B}  \textbf{82}, 161414(R) (2010).
	
\bibitem{QAHE-4}
K. F. Garrity and D. Vanderbilt, Chern Insulators from Heavy Atoms on Magnetic Substrates, \text{Phys. Rev. Lett.}  \textbf{110}, 116802 (2013).
	
\bibitem{QAHE-5}
C. Fang, M. J. Gilbert, and B. A. Bernevig, Large-Chern-Number Quantum Anomalous Hall Effect in Thin-Film Topological Crystalline Insulators, \text{Phys. Rev. Lett.}  \textbf{112}, 046801 (2014).
	
\bibitem{QAHE-6}
Y. Ren, J. Zeng, X. Deng, F. Yang, H. Pan, and Z. Qiao, Quantum anomalous Hall effect in atomic crystal layers from in-plane magnetization, \text{Phys. Rev. B}  \textbf{94}, 085411 (2016).
	
\bibitem{QAHE-7}
H. P. Wang, W. Luo, and H. J. Xiang, Prediction of high-temperature quantum anomalous Hall effect in two-dimensional transition-metal oxides, \text{Phys. Rev. B}  \textbf{95}, 125430 (2017).
	
\bibitem{QAHE-8}
M. M. Otrokov, I. P. Rusinov, M. Blanco-Rey, M. Hoffmann, A. Y. Vyazovskaya, S. V. Eremeev, A. Ernst, P. M. Echenique, A. Arnau, and E. V. Chulkov, Unique Thickness-Dependent Properties of the van der Waals Interlayer Antiferromagnet MnBi$_2$Te$_4$ Films, \text{Phys. Rev. Lett.}  \textbf{122}, 107202 (2019).
	
\bibitem{QAHE-9}
F. Wu, T. Lovorn, E. Tutuc, I. Martin, and A. H. MacDonald, Topological Insulators in Twisted Transition Metal Dichalcogenide Homobilayers, \text{Phys. Rev. Lett.} \textbf{122}, 086402 (2019).
	
\bibitem{QAHE-10}
Y.-H. Zhang, D. Mao, Y. Cao, P. Jarillo-Herrero, and T. Senthil, Nearly flat Chern bands in moiré superlattices, \text{Phys. Rev. B}  \textbf{99}, 075127 (2019).
	
\bibitem{QAHE-11}
N. Bultinck, S. Chatterjee, and M. P. Zaletel, Mechanism for Anomalous Hall Ferromagnetism in Twisted Bilayer Graphene, \text{Phys. Rev. Lett.}  \textbf{124}, 166601 (2020).
	
\bibitem{QAHE-12}
J. Shi, J. Zhu, and A. H. MacDonald, Moir$\acute{\text{e}}$ commensurability and the quantum anomalous Hall effect in twisted bilayer graphene on hexagonal boron nitride, \text{Phys. Rev. Lett.}  \textbf{103}, 075122 (2021).
	
	
\bibitem{QAHE-13}
G. Chen, A. L. Sharpe, E. J. Fox, Y.-H. Zhang, S. Wang, L. Jiang, B. Lyu, H. Li, K. Watanabe, T. Taniguchi et al., and F. Wang, Tunable correlated Chern insulator and ferromagnetism in a moiré superlattice, \text{Nature}  \textbf{579}, 56 (2020).
	
\bibitem{QAHE-14}
Z. Li, Y. Han, and Z. Qiao , Chern Number Tunable Quantum Anomalous Hall Effect in Monolayer Transitional Metal Oxides via Manipulating Magnetization Orientation, \text{Phys. Rev. Lett.}  \textbf{129}, 036801 (2022).
	
\bibitem{liufeng}
Feng Liu, Two-dimensional topological insulators: past, present, and future, Coshare Science,  \textbf{01}, v3, 1-62 (2023).	
	
\bibitem{QAHE-15}
C. Z. Chang, J. S. Zhang, X. Feng, J. Shen, Z. C. Zhang, M. H. Guo, K. Li, Y. B. Ou, P. Wei, L. L. Wang, et al., Experimental Observation of the Quantum Anomalous Hall Effect in a Magnetic Topological Insulator, \text{Science}  \textbf{340}, 167 (2013).
	
\bibitem{QAHE-16}
Y. Deng, Y. Yu, M. Z. Shi, Z. Guo, Z. Xu, J. Wang, X. H. Chen, and Y. Zhang, Quantum anomalous Hall effect in intrinsic magnetic topological insulator MnBi$_2$Te$_4$, \text{Science}  \textbf{367}, 895 (2020).
	
\bibitem{QAHE-17}
M. Serlin, C. L. Tschirhart, H. Polshyn, Y. Zhang, J. Zhu, K. Watanabe, T. Taniguchi, L. Balents, and A. F. Young, Intrinsic quantized anomalous Hall effect in a moiré heterostructure, \text{Science}  \textbf{367}, 900 (2020).
	
\bibitem{QAHE-18}
T. Li, S. Jiang, B. Shen, Y. Zhang, L. Li, Z. Tao, T. Devakul, K. Watanabe, T. Taniguchi, L. Fu et al., Quantum anomalous Hall effect from intertwined moiré bands, \text{Nature}  \textbf{600}, 641 (2021).
	
	
	
\bibitem{QAHE-19}
C.-X. Liu, S.-C. Zhang, and X.-L. Qi, The Quantum Anomalous Hall Effect: Theory and Experiment, \text{Annu. Rev. Condens. Matter Phys.}  \textbf{7}, 301 (2016).
	
\bibitem{AFM-review-1}
V. Baltz, A. Manchon, M. Tsoi, T. Moriyama, T. Ono, and Y. Tserkovnyak, Antiferromagnetic spintronics, \text{Rev. Mod. Phys.} \textbf{90}, 015005 (2018).
	
\bibitem{AFM-review-2}
T. Jungwirth, X. Marti, P. Wadley, and J. Wunderlich, Antiferromagnetic spintronics, \text{Nat. Nanotechnol.} \textbf{11}, 231 (2016).
	
	
\bibitem{AFM-review-3}
L. Šmejkal, Y. Mokrousov, B. Yan and A. H. MacDonald, Topological antiferromagnetic spintronics, \text{Nat. Phys.} \textbf{14}, 242 (2018).
	
\bibitem{AFM-review-4}
P. Němec, M. Fiebig, T. Kampfrath and A. V. Kimel, Antiferromagnetic opto-spintronics, \text{Nat. Phys.} \textbf{14}, 229 (2018).
	
\bibitem{AFM-review-5}
S. Rahman, J. F. Torres, A. R. Khan, and Y. Lu, Recent Developments in van der Waals Antiferromagnetic 2D Materials: Synthesis, Characterization, and Device Implementation, \text{ACS Nano} \textbf{15}, 17175 (2021).
	
\bibitem{AFM-1}
B.Huang, G. Clark, E. Navarro-Moratalla, D. R. Klein, R. Cheng, K. L. Seyler, D. Zhong, E. Schmidgall, M. A. McGuire, D. H. Cobden et al., Layer-dependent ferromagnetism in a van der Waals crystal down to the monolayer limit, \text{Nature} \textbf{546}, 270 (2017).
	
\bibitem{AFM-2}
B. Huang, G. Clark, D. R. Klein, D. MacNeill, E. Navarro-Moratalla, K. L. Seyler, N. Wilson, M. A. McGuire, D. H. Cobden, D. Xiao, et al., Electrical control of 2D magnetism in bilayer CrI$_3$, \text{Nat. Nanotech.} \textbf{13}, 544 (2018)
	
\bibitem{AFM-3}
W. Chen, Z. Sun, Z. Wang, L. Gu, X. Xu, S. Wu, and C. Gao, Direct observation of van der Waals stacking–dependent interlayer magnetism, \text{Science} \textbf{366}, 983 (2019).
	
\bibitem{AFM-4}
Z. Sun, Y. Yi, T. Song, G. Clark, B. Huang, Y. Shan, S. Wu, D. Huang, C. Gao, Z. Chen et al., Giant nonreciprocal second-harmonic generation from antiferromagnetic bilayer CrI$_3$, \text{Nature} \textbf{572}, 497 (2019).
	
\bibitem{AFM-QAHE-1}
P. Zhou, C. Q. Sun, and L. Z. Sun, Two Dimensional Antiferromagnetic Chern Insulator: NiRuCl$_6$, \text{Nano. Lett.} \textbf{16}, 6325 (2016). 
	
\bibitem{AFM-QAHE-2}
J. Wang, Antiferromagnetic Dirac semimetals in two dimensions, \text{Phys. Rev. B} \textbf{95}, 115138 (2017) 
	
\bibitem{AFM-QAHE-3}
K. Jiang, S. Zhou, X. Dai, and Z. Wang, Antiferromagnetic Chern Insulators in Noncentrosymmetric Systems, \text{Phys. Rev. Lett.} \textbf{120}, 157205 (2018)
	
\bibitem{AFM-QAHE-4}
X. Li, A. H. MacDonald, and H. Chen, Quantum Anomalous Hall Effect through Canted Antiferromagnetism, arXiv:1902.10650. 
	
\bibitem{AFM-QAHE-5}
P.-J. Guo, Z.-X. Liu  and Z.-Y. Lu, Quantum anomalous hall effect in collinear antiferromagnetism, \text{npj Comput. Mater.} \textbf{9}, 70 (2023)
	
\bibitem{AFM-QAHE-6}
B. Wu , Y.-L. Song, W.-X. Ji, P.-J. Wang, S.-F. Zhang, and C.-W. Zhang, Quantum anomalous Hall effect in an antiferromagnetic monolayer of MoO, \text{Phys. Rev. B} \textbf{107}, 214419 (2023)
	
\bibitem{AFM-QAHE-7}
M. Ebrahimkhas, G. S. Uhrig, W. Hofstetter, and M. Hafez-Torbati, Antiferromagnetic Chern insulator in centrosymmetric systems, \text{Phys. Rev. B} \textbf{106}, 205107 (2022)
	
\bibitem{AFM-QAHE-8}
S. Du, P. Tang, J. Li, Z. Lin, Y.Xu, W. Duan, and A. Rubio, Berry curvature engineering by gating two-dimensional antiferromagnets, \text{Phys. Rev. Research} \textbf{2}, 022025(R) (2020)

\bibitem{AFM-QAHE-Lei}
C. Lei, T. V. Trevisan, O. Heinonen, R. J. McQueeney, and A. H. MacDonald, Quantum anomalous Hall effect in perfectly compensated collinear antiferromagnetic thin films, \text{Phys. Rev. B} \textbf{106}, 195433 (2022)
	
\bibitem{AFM-QAHE-9}
X. Zhou, W. Feng, Y. Li, and Y. Yao, Spin-Chirality-Driven Quantum Anomalous and Quantum Topological Hall Effects in Chiral Magnets, \text{Nano Lett.} \textbf{23}, 5680-5687 (2023)
	
\bibitem{AFM-QAHE-10}
Y. Liu, J. Li and Q. Liu, Chern-Insulator Phase in Antiferromagnets, \text{Nano Lett.} \textbf{23}, 8650-8656 (2023)
	

	
\bibitem{MBT-0}
D. Ovchinnikov, X. Huang, Z. Lin, Z. Fei, J. Cai, T. Song, M. He,
Q. Jiang, C. Wang, H. Li et al.,Intertwined Topological and Magnetic Orders in Atomically ThinChern Insulator MnBi$_2$Te$_4$, \text{Nano Lett.} \textbf{21}, 2544 (2021)

\bibitem{MBT-01}
Y.-J. Hao, P. Liu, Y. Feng, X.-M. Ma, E. F. Schwier, M. Arita, S. Kumar, C. Hu, R. Lu, M. Zeng et al., Gapless Surface Dirac Cone in Antiferromagnetic Topological Insulator MnBi$_2$Te$_4$, \text{Phys. Rev. X} \textbf{9}, 041038 (2019)
	
\bibitem{MBT-1}
M. M. Otrokov, I. I. Klimovskikh, H. Bentmann, D. Estyunin, A. Zeugner, Z. S. Aliev, S. Gaß, A. U. B. Wolter, A. V. Koroleva, A. M. Shikin et al., Prediction and observation of an antiferromagnetic topological insulator, \text{Nature} \textbf{576}, 416 (2019)
	
\bibitem{MBT-2}
J. Li, C. Wang, Z. Zhang, B.-L. Gu, W. Duan, and Y. Xu, Magnetically controllable topological quantum phase transitions in the antiferromagnetic topological insulator MnBi$_2$Te$_4$, \text{Phys. Rev. B} \textbf{100}, 121103(R) (2019)
	
\bibitem{MBT-3}
W. Liang, J. Zeng, Z. Qiao, Y. Gao, and Q. Niu, Berry-Curvature Engineering for Nonreciprocal Directional Dichroism in Two-Dimensional Antiferromagnets, \text{Phys. Rev. Lett.} \textbf{131}, 256901 (2023).
	
\bibitem{MBT-4}
J. Li, Y. Li, S. Du, Z. Wang, B. L. Gu, S. C. Zhang, K. He, W. Duan, and Y. Xu, Intrinsic magnetic topological insulators in van der Waals layered MnBi$_2$Te$_4$-family materials, \text{Sci. Adv.} \textbf{5}, eaaw5685 (2019).
	
\bibitem{MBT-5}
J. Ge, Y. Liu, J. Li, H. Li, T. Luo, Y. Wu, Y. Xu, and J. Wang, High-Chern-number and high-temperature quantum Hall effect without Landau levels, \text{Natl. Sci. Rev.} \textbf{7}, 1280 (2020).

\bibitem{layer Hall} 
A. Gao, Y.-F. Liu, C. Hu, J.-X. Qiu, C. Tzschaschel, B. Ghosh, S.-C. Ho, D. Bérubé, R. Chen, H. Sun et al., Layer Hall effect in a 2D topological axion antiferromagnet, \text{Nature}  \textbf{595}, 521 (2021).

\bibitem{MBT-6}
P. M. Sass, W. Ge, J. Yan, D. Obeysekera, J. J. Yang, and W. Wu, Magnetic Imaging of Domain Walls in the Antiferromagnetic Topological Insulator MnBi$_2$Te$_4$, \text{Nano Lett.} \textbf{20}, 2609-2614 (2020).
	
\bibitem{MBT-7}
D. Zhang, M. Shi, T. Zhu, D. Xing, H. Zhang, and J. Wang, Topological Axion States in the Magnetic Insulator MnBi$_2$Te$_4$ with the Quantized Magnetoelectric Effect, \text{Phys. Rev. Lett.} \textbf{122}, 206401 (2019).
	
\bibitem{MBT-8}
H. Sun, B. Xia, Z. Chen, Y. Zhang, P. Liu, Q. Yao, H. Tang, Y. Zhao, H. Xu, and Q. Liu, Rational Design Principles of the Quantum Anomalous Hall Effect in Superlatticelike Magnetic Topological Insulators, \text{Phys. Rev. Lett.} \textbf{123}, 096401 (2019).
	
\bibitem{MBT-9}
H.-P. Sun, C. M. Wang, S.-B. Zhang, R. Chen, Y. Zhao, C. Liu, Q. Liu, C. Chen, H.-Z. Lu, and X. C. Xie, Analytical solution for the surface states of the antiferromagnetic topological insulator MnBi$_2$Te$_4$, \text{Phys. Rev. B} \textbf{102}, 241406(R) (2020).
	
\bibitem{MBT-10}
W. Liang, T. Hou, J. Zeng, Z. Liu, Y. Han, and Z. Qiao, Layer-dependent zero-line modes in antiferromagnetic topological insulators, \text{Phys. Rev. B}  \textbf{107}, 075303 (2023).
	
\bibitem{MBT-11}
B. Lian, Z. Liu, Y. Zhang, and J. Wang, Flat Chern Band from Twisted Bilayer MnBi$_2$Te$_4$, \text{Phys. Rev. Lett.}  \textbf{124}, 126402 (2020).



\bibitem{exp-1}
Y. Li, Y. Bai, Y. Feng, J. Luan, Z. Gao, Y. Chen, Y. Tong, R. Liu, S. K. Chong, K. L. Wang et al., Reentrant quantum anomalous Hall effect in molecular beam epitaxy-grown MnBi$_2$Te$_4$ thin films, arXiv:2401.11450.	

\bibitem{exp-2}
Z. Lian, Y. Wang, Y. Wang, Y.Feng, Z. Dong, S. Yang, L. Xu, Y. Li, B. Fu, Y. Li et al., Antiferromagnetic Quantum Anomalous Hall Effect Modulated by Spin Flips and Flops, arXiv:2405.08686.		

\bibitem{exp-3}
Y. Wang, B. Fu, Y. Wang, Z. Lian, S. Yang, Y.Li, L. Xu, Z. Gao, W. Jiang, J. Zhang et al., Towards the Quantized Anomalous Hall effect in AlO$_x$-capped MnBi$_2$Te$_4$, arXiv:2405.08677.		



	
\bibitem{jianghua}
H. Jiang, Z. Qiao, H. Liu, and Q. Niu, Quantum anomalous Hall effect with tunable Chern number in magnetic topological insulator film, \text{Phys. Rev. B}  \textbf{85}, 045445 (2012).
	
\bibitem{BHZ}
H. Zhang, C.-X. Liu, X.-L. Qi, X. Dai, Z. Fang, and S.-C. Zhang, Topological insulators in Bi2Se3, Bi2Te3 and Sb2Te3 with a single Dirac cone on the surface, \text{Nat. Phys.}  \textbf{5}, 438 (2009).

\bibitem{chengran}
Y.-H. Li and R. Cheng, Identifying axion insulator by quantized magnetoelectric effect in antiferromagnetic MnBi$_2$Te$_4$ tunnel junction, \text{Phys. Rev. Res.}  \textbf{4}, L022067 (2022).


\bibitem{doping}
N. Ehlen, Y. Falke, B. V. Senkovskiy, T. Knispel, J. Fischer, O. N. Gallego, C. Tresca, M. Buchta, K. P. Shchukin, Alessandro D'Elia et al., Orbital-selective chemical functionalization of MoS$_2$ by Fe, \text{Phys. Rev. B} \textbf{108}, 195430 (2023)	

\bibitem{pressure}
L. Craco and S. Leoni, Pressure-induced orbital-selective metal from the Mott insulator BaFe$_2$Se$_3$, \text{Phys. Rev. B} \textbf{101}, 245133 (2020)	

	
\bibitem{supple} See Supplemental Materials, which includes Ref.~\cite{QAHE-8,Landauer-Buttiker,vasp,GGA,MP,GGA-U,wannier90,wannier90-QAHE,wannier_tools,ybh-sa}.


\bibitem{Efield-1}
R. Mei, Y.-F. Zhao, C. Wang, Y. Ren, D. Xiao, C.-Z. Chang, and C.-X. Liu, Electrically Controlled Anomalous Hall Effect and Orbital Magnetization in Topological Magnet MnBi$_2$Te$_4$, \text{Phys. Rev. Lett.} \textbf{132}, 066604 (2024)

\bibitem{Efield-2}
R. Mei, D. Kaplan, B. Yan, C.-Z. Chang, and C.-X. Liu, Electrical control of intrinsic nonlinear Hall effect in antiferromagnetic topological insulator sandwiches, arXiv:2406.02738


\bibitem{ChernMetal}
A. M. Cook and A. Paramekanti, Double Perovskite Heterostructures: Magnetism, Chern Bands, and Chern Insulators, \text{Phys. Rev. Lett.}  \textbf{113}, 077203 (2014).


\bibitem{bulk-boundary}
R. S. K. Mong and V. Shivamoggi, Edge states and the bulk-boundary correspondence in Dirac Hamiltonians, \text{Phys. Rev. B} \textbf{83}, 125109 (2011)	
	
\bibitem{pinning}
E. Liu, Y.-C. Wu, S. Couet, S. Mertens, S. Rao, W. Kim, K. Garello, D. Crotti, S. Van Elshocht, J. De Boeck, G. S. Kar, and J. Swerts, Synthetic-Ferromagnet Pinning Layers Enabling Top-Pinned Magnetic Tunnel Junctions for High-Density Embedded Magnetic Random-Access Memory, \text{Phys. Rev. Appl.} \textbf{10}, 054054 (2018)

\bibitem{pinning-1}
B. Chen, X. Liu, Y.-H. Li, H. Tay, T. Taniguchi, K. Watanabe, M. H. W. Chan, J. Yan, F. Song, R. Cheng et al., Even-Odd Layer-Dependent Exchange Bias Effect in MnBi$_2$Te$_4$ Chern Insulator Devices, arXiv:2404.03032

\bibitem{pinning-2}
S. K. Chong, Y. Cheng, H. Man, S. H. Lee, Y. Wang, B.Dai, M. Tanabe, T.H. Yang, Z. Mao, K. A. Moler et al., Intrinsic exchange biased anomalous Hall effect in an uncompensated antiferromagnet MnBi$_2$Te$_4$, \text{Nat. Commun.} \textbf{15}, 2881 (2024)

\bibitem{pinning-3}
H. Fu, C.-X. Liu, and B. Yan, Exchange bias and quantum anomalous Hall effect in the MnBi$_2$Te$_4$/CrI$_3$ heterostructure, \text{Sci. Adv.} \textbf{6}, eaaz0948(2020)

\bibitem{pinning-4}
J.-Z. Fang, H.-N. Cui, S. Wang, J.-D. Lu, G.-Y. Zhu, X.-J. Liu, M.-S. Qin, J.-K. Wang, Z.-N. Wu, Y.-F. Wu et al., Exchange bias in the van der Waals heterostructure $\rm{MnBi_2Te_4/Cr_2Ge_2Te_6}$, \text{Phys. Rev. B} \textbf{107}, L041107(2023)

\bibitem{heisenberg}
Y. Lai, L. Ke, J. Yan, R. D. McDonald, and R. J. McQueeney, Defect-driven ferrimagnetism and hidden magnetization in MnBi$_2$Te$_4$, \text{Phys. Rev. B} \textbf{103}, 184429 (2021)





\bibitem{Landauer-Buttiker}
S. Datta, Electronic Transport in Mesoscopic Systems (Cambridge University Press, Cambridge, 1997).	



%DFT
\bibitem{vasp} G. Kresse and J. Furthm{\"u}ller, Efficient iterative schemes for ab initio total-energy calculations using a plane-wave basis set, Phys. Rev. B \textbf{54}, 11169 (1996).%

\bibitem{GGA} J. P. Perdew, K. Burke, and M. Ernzerhof, Generalized Gradient Approximation Made Simple, Phys. Rev. Lett. \textbf{77}, 3865 (1996).%

\bibitem{MP} H. J. Monkhorst and J. D. Pack, Special points for Brillouin-zone integrations, Phys. Rev. B \textbf{13}, 5188 (1976).%

\bibitem{GGA-U} L. Wang, T. Maxisch, and G. Ceder, Oxidation energies of transition metal oxides within the GGA+U framework, Phys. Rev. B \textbf{73}, 195107 (2006).%

\bibitem{wannier90} A. A. Mostofi, J. R. Yates, Y.-S. Lee, I. Souza, D. Vanderbilt, and N. Marzari, wannier90: A tool for obtaining maximally-localised Wannier functions, Comput. Phys. Commun. \textbf{178}, 685 (2008).%

\bibitem{wannier90-QAHE} X. Wang, J. R. Yates, I. Souza, and D. Vanderbilt, Ab initio calculation of the anomalous Hall conductivity by Wannier interpolation, Phys. Rev. B \textbf{74}, 195118 (2006).%

\bibitem{wannier_tools} Q. S. Wu, S. N. Zhang, H.-F. Song, M. Troyer, and A. A. Soluyanov, WannierTools: An open-source software package for novel topological materials, Comput. Phys. Commun. \textbf{224}, 405 (2018).%

\bibitem{ybh-sa} H. Fu, C.-X. liu and B. Yan, Exchange bias and quantum anomalous Hall effect in the MnBi$_{2}$Te$_{4}$/CrI$_{3}$ heterostructure, Sci. Adv. \textbf{6}, eaaz0948 (2020).

	
\end{thebibliography}
\end{document}